\documentclass[11pt,twoside]{article}

\usepackage{asp2004}
\usepackage{epsf}

\begin{document}

\title{Mg II Absorbing Galaxies: Morphologies and Kinematics}

\author{Chris Churchill$^{1}$, Chuck Steidel$^{2}$, and Glenn Kacprzak$^{1}$}

\affil{$^{1}$ Department of Astronomy, New Mexico State University, 
MSC 4500, P.O. Box 30001, Las Cruces, NM, 88003} 

\affil{$^{2}$ Department of Astronomy, Caltech, MS 105-24, Pasadena, CA 91125}

\begin{abstract}
In this contribution, we review our current knowledge of the
properties of galaxies, and their extended halos, selected by
\hbox{Mg~{\sc ii}} absorption in the spectra of background quasars.
We then describe recent efforts to quantify the morphologies and
orientations of galaxies and explore how these relate to the gas
kinematics.  In a sample of 26 galaxies, we find no clear connection
between the orientation of the quasar line of sight through the galaxy
and the velocity spread of the gas.  However, it appears that the
quantity of gas ``stirred up'' in the halo may be correlated to
asymmetry in the galaxy morphology. Since the galaxies have fairly
normal morphologies, this connection may suggest that galaxies with
extended halos experienced an interaction or merging event a few
dynamical times prior to the epoch of observation.
\end{abstract}

\section{Quasar Absorption Lines: Further and Farther}

From the study of the Galaxy and local galaxies, the disk/halo
interface is now known to be a dynamic region, probably built in
response to star forming regions and the infall of ``high velocity
clouds'' (HVCs).  Using sensitive 21--cm emission line imaging of
local galaxies, several contributers of this meeting have reported
himneys, fountains, superbubbles, and \hbox{H~{\sc i}} holes (see
contributions throughout this volume).  Extending somewhat further
into the halo are the so-called ``\hbox{H~{\sc i}} beards''
\citep{sancisi01} and ``forbidden gas'' complexes
\citep{fraternali02}.  Beards are cold neutral gas structures several
kpc from the disks that depart from galaxy rotation by as much as
$\sim 50$--100~km~s$^{-1}$; they are suggestive of a ``lagging halo".
Forbidden gas structures have kinematics that run counter to the
galaxy.  The diffuse interstellar gas (DIGs) in local galaxies is seen
out to $\sim 10$~kpc above the galactic disks
\cite[e.g.,][]{swater97,rand00}.  DIG gas often exhibits decreasing
rotational velocities with height above the disk (i.e., ``halo lag'').
Simple fountain flow models are consistent with this behavior
\citep{collins02}.

In order to make greater sense of these observations and ultimately of
the role of gas in galaxy evolution, it would be desirable to build an
observational bridge connecting slightly earlier cosmic epochs to the
present.  It would also be desirable to probe further out into the
galactic halos in order to firmly establish the nature of gas far from
galaxy disks.  The technique of quasar absorption lines is uniquely
suited to the task; it provides a natural window on the gaseous
conditions in galaxies over a broad range of redshifts (cosmic epochs)
and allows for the study of {\it both\/} the inner and outer regions
of galaxies and their halos.

\section{Mg II Absorption: Case for the Ideal Probe}

The strong resonant \hbox{Mg~{\sc ii}} $\lambda\lambda 2796, 2803$
doublet in absorption is quite arguably the best tracer of
metal-enriched gas {\it associated with galaxies}.  \hbox{Mg~{\sc ii}}
absorption is common and easily spotted in quasar spectra.
Furthermore, magnesium, an $\alpha$--process element, is ejected by
supernovae into interstellar, intergroup, and intergalactic space from
the epoch of the very first generation of stars at the highest
redshifts. Since \hbox{Mg~{\sc ii}} absorption is known to arise in
gas spanning five decades of neutral hydrogen column density, from
$\log N(\hbox{H~{\sc i}}) = 15.5$ to $20.5$~atoms~cm$^{-2}$
\citep[e.g.,][]{ber86,weakI}, it samples a large range of
astrophysical environments near star forming and post--star forming
regions (i.e., galaxies).

\section{Absorption--Selected Galaxies}

Following the initial study by \citet{bb91}, \citet[][hereafter
SDP]{sdp94}, obtained images of roughly 50 quasar fields and
identified the ``absorption--selected galaxies'' associated with
strong \hbox{Mg~{\sc ii}} systems.  Their study firmly established a
connection between the galaxy luminosities and colors and the extent
of their gaseous ``\hbox{Mg~{\sc ii}} absorbing envelopes'' for $0.3
\leq z \leq 1.0$

\begin{figure}[htb]
\plotfiddle{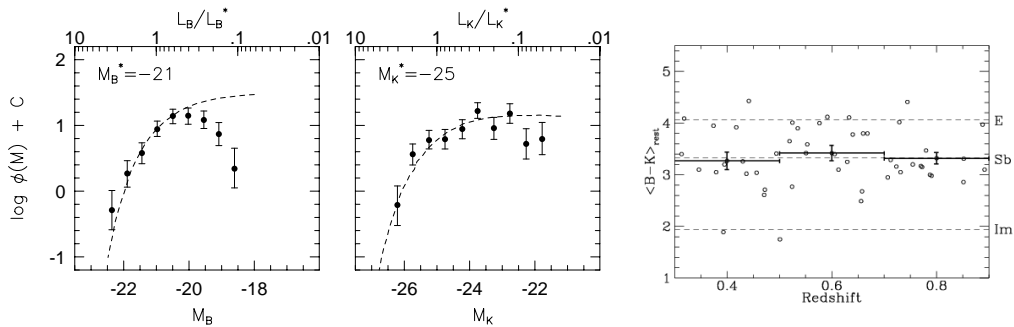}{1.4in}{0}{65.}{65.}{-196}{-10}
\caption {{\bf Global Galaxy Properties} --- (left, center) The
B--band and K--band luminosity functions of Mg~{\sc ii} absorption
selected galaxies.  The dashed lines are the luminosity functions of
$z \sim 0$ field galaxies overlayed on the data.  There is a paucity
of faint blue galaxies relative to the present day distribution of
field galaxies. --- (right) The rest--frame $B-K$ colors of Mg~{\sc
ii} absorption selected galaxies.  The average color is consistent
with that of an Sb spiral.}
\label{fig:BKlum}
\end{figure}

As shown in the left and center panels of Figure~\ref{fig:BKlum}, the
luminosities of \hbox{Mg~{\sc ii}} selected galaxies exhibit
little evolution with redshift.  However, there is differential
evolution in the luminosity function
\citep[see][]{lilly96,guillemin97} in that faint blue galaxies are
apparently under abundant; they are not selected in a survey using
\hbox{Mg~{\sc ii}} gas cross section, seen as a turn down at the
faint--end of the B--band luminosity.  This is a quandary in that low
mass halos of star forming galaxies are most expected to have large
gas cross sections (as will be discussed below, this may be a
selection bias).

\begin{figure}[htb]
\plotfiddle{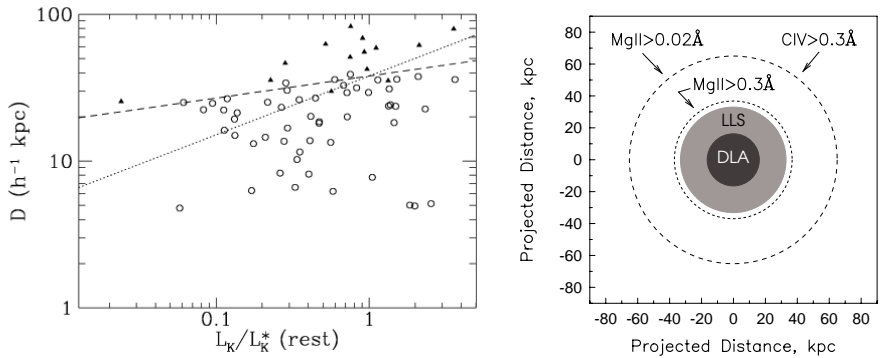}{1.7in}{0}{65.}{65.}{-186}{-3}
\caption {{\bf Galaxy Gas Cross Section} --- (left) Impact parameter,
$D$, versus galaxy luminosity (Steidel 1995). The extent of the
absorbing gas has a weak dependence upon galaxy luminosity, $R \propto
L^{0.15}$, as shown by the dashed line.  Open circles are absorbing
galaxies and solid triangles are non--absorbing galaxies.  --- (right)
The inferred statistical cross sections of Mg~{\sc ii} absorbers
assuming they are associated with normal bright galaxies.}
\label{fig:RLK}
\vglue -0.2in
\end{figure}

\citet{steidel95} showed that the sizes of \hbox{Mg~{\sc ii}}
``halos'' scale weakly with galaxy luminosity, following, $R(L_K)
\simeq 38h^{-1} \left(L_K / L^{\ast}_K\right)^{0.15}$~kpc. This
relationship, illustrated in the left panel of Figure~\ref{fig:RLK},
was determined by plotting the impact parameter as a function of host
galaxy luminosity and maximizing the number of absorbing galaxies
(open circles) below the $R(L_K)$ relation and the number of
non--absorbing galaxies (solid triangles) above the $R(L_K)$ relation.
From the clean demarcation (as marked by the dashed line) it was
inferred that the halos are roughly spherical and have unity covering
factors and sharp boundaries for $W(\hbox{Mg~{\sc ii}}) > 0.3$~{\AA}
(the dotted line is the canonical Holmberg slope 0.4, normalized at $D
= 38h^{-1}$ kpc).

As illustrated in the right panel of Figure~\ref{fig:RLK}, the gas
cross section for $W(\hbox{Mg~{\sc ii}}) > 0.3$~{\AA} is $\sim 40$~kpc
and is statistically consistent the extent of gas optically thick at
the Lyman limit of neutral hydrogen (i.e., LLS).  However, for the
so--called weak \hbox{Mg~{\sc ii}} absorbers, those with
$W(\hbox{Mg~{\sc ii}}) <0.3$~{\AA} \citep{weakI}, little is known
about what galaxy types are selected.  If it is assumed that normal
bright galaxies give rise to weak \hbox{Mg~{\sc ii}} absorption, then,
using the normalization appropriate for the luminosity functions
presented in Figure~{\ref{fig:BKlum}, their gas cross section is $\sim
70$~kpc and is consistent with the region occupied by strong
\hbox{C~{\sc iv}} absorbing gas.

\section{Need for Complete Census}

For all that has been accomplished to date, it is probable the
scenario of a spherical distribution of gas with unity covering factor
may be oversimplified, or a result of survey methodology.  Indeed, the
SDP survey remains relatively imcomplete in that 30\% of the absorbing
galaxies in the fields do \underbar{not} have spectroscopically
measured redshifts (i.e., are \underbar{not confirmed} to be at the
absorption redshift).

Though 25 ``control'' fields, in which none of the galaxies gave rise
to strong \hbox{Mg~{\sc ii}} absorption, were systematically studied
(i.e., all galaxies having $L>0.05L^{\ast}$ within 10{\arcsec} have
measured photometry and spectroscopic redshifts), this was not the
case for the 52 fields with absorbing galaxies.  In these latter
fields, the spectroscopic redshifts were not completed out to a fixed
radius around the quasars; once a galaxy was identified at the
absorber redshift, no further redshifts were determined in the field.
The galaxy spectroscopy was performed in an outward pattern from the
quasar and galaxies that were targeted were based upon their
magnitudes and colors being consistent with the absorber redshift.
Close in galaxies of unexpected luminosity and color and further out
galaxies could have been missed as potential absorbers
\citep[see][]{bbp95}.

Thus, there is the very real possibility that some galaxies have been
misidentified as the absorber, or their may be more than one galaxy
giving rise to the \hbox{Mg~{\sc ii}} absorption.  Such incompleteness
in the survey could have strong implications on our inference of the
covering factor, luminosity functions, and geometric model of the
absorbing gas halos.  \citet{cc96} studied the effects of
incompleteness and possible misidentifications in the SDP survey and
concluded that the realization of absorbing/non-absorbing galaxies
shown in Figure~\ref{fig:RLK} obtained by Steidel (1995) was not
improbable for the survey methods even if the covering factor was as
small as 70\%.

For only a single quasar field, that of 3C~336 has a thorough census
of absorbing and non--absorbing galaxies been completed
\citep{steidel97}.  This rich field had four known strong
\hbox{Mg~{\sc ii}} absorbers, but in the course of the study two
additional weaker systems were found with $W(\hbox{Mg~{\sc ii}}) <
0.3$~{\AA}!  A deep WFPC--2/{\it HST\/} image of the field was
obtained and every galaxy within 50{\arcsec} of the quasar in the
field with $L>0.05L_{K}^{\ast}$ was spectroscopically redshifted.

A Damped \hbox{Ly$\alpha$} absorber (DLA), still remains unidentified.
Interestingly, there is a galaxy at the DLA redshift; if it is the
absorbing galaxy it lies well {\it outside\/} the $R(L_K)$ relation
($L=0.6L_{K}^{\ast}$ and $D=65h^{-1}$~kpc) and has an extraordinarly
large $N(\hbox{H~{\sc i}})$ for this impact parameter.

Furthermore, at least one non--absorbing galaxy was identified that is
well below the $R(L_K)$ relation so that it would be predicted to give
rise to strong \hbox{Mg~{\sc ii}}.  An after--the--fact search for
\hbox{Mg~{\sc ii}} absorbtion in the HIRES/Keck spectrum was
inconclusive; this galaxy {\it may\/} host very weak \hbox{Mg~{\sc
ii}} absorber.  For each of two additional cases, there are two
galaxies having redshifts coincident with the \hbox{Mg~{\sc ii}}
absorption redshift.

Over the last decade, several quasar fields have been reexamined by us
in a patch work effort and a disturbing number of misidentifications
and multiple galaxies at the absorption redshift have indeed been
found!  For example, in one case, a galaxy at 70$h^{-1}$~kpc was found
to be the absorber, not the assumed galaxy at 30$h^{-1}$~kpc.  This
case again demonstrates that some of the absorbing galaxies lie well
above the $R(L_K)$ relation (see Figure~\ref{fig:RLK}).

In another case, the spectroscopic redshift of a galaxy thought to be
associated with a strong \hbox{Mg~{\sc ii}} absorber revealed it to be
at a redshift with no known absorption .  When reexamined with higher
senstivity quasar spectra, the galaxy was found to align with a weak
\hbox{Mg~{\sc ii}} absorber at a different redshift; there is no
obvious galaxy candidate in the field for the strong absorber!  In the
SDP survey, the galaxy associated with weak \hbox{Mg~{\sc ii}}, if
properly identified, would have been a non--absorber below the
$R(L_K)$ relation.

In yet another field, a double galaxy was found at the absorber
redshift; the assumed galaxy was not the absorber.  Their are at least
two other fields in which three or more galaxies are clustered at the
absorption redshift.

Clearly, our recent work and the complete analysis of the 3C~336
field has provided examples of departures from the conclusions drawn
by SDP.  What is also very important to realize is that the HIRES/Keck
quasar spectroscopy that has uncovered weak \hbox{Mg~{\sc ii}}
absorption has shed new levels of sensitivity and complications on the
whole \hbox{Mg~{\sc ii}} halo gas--galaxy connection.  Some galaxies
show weak absorption at smaller galactocentric distances and some show
it at larger galactocentric distances.  Armed with a factor of 10
increase in sensitivity to \hbox{Mg~{\sc ii}} absorption using
HIRES/Keck, we now have the potential leverage to probe the patchiness
of the halos and measure the covering factor at much more sensitive
column density thresholds.  All this, of course, hinges on the
paramount objective of obtaining an unbiased, accurate, and complete
census of the galaxies.

\section{Galaxy Morphologies and Gas Kinematics}

There has been a fair amount of activity in examining the kinematics
of $z \leq 1$ \hbox{Mg~{\sc ii}} absorbers and comparing them with the
{\it gross\/} properties of the absorbing galaxies
\citep{lb92,csv96,archiveII,cv01}.  However, knowledge of the galaxy
has been limited to the photometric magnitudes, $B-K$ colors, and the
observed impact parameters.

\begin{figure}[bht]
\plotfiddle{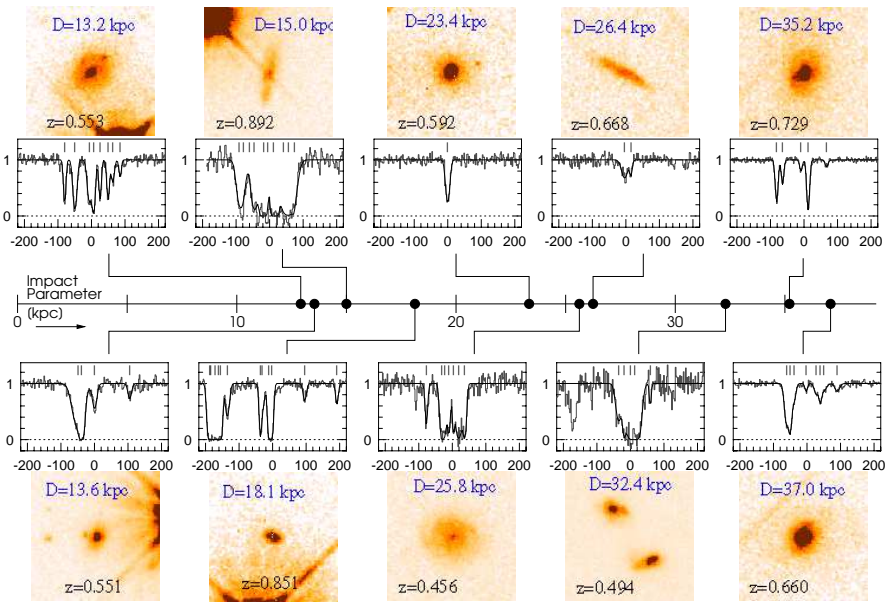}{2.90in}{-90.}{45.}{45.}{-170.}{225.}
\caption{{\bf Morphology, Impact Parameter, and Gas Kinematics} ---
The galaxy morphology (from WFPC-2/{\it HST\/} images) and the
Mg~{\sc ii} $\lambda 2796$ line--of--sight kinematics (from HIRES/Keck
quasar spectra) shown with increasing impact parameter order for 10 of
our 26 galaxies.  The vertical bars above the Mg~{\sc ii}~$\lambda
2796$ profiles (in velocity) mark the number of subcomponents and
their velocities based upon Voigt profile fitting.}
\label{fig:collage}
\end{figure}

The sample size has remained relatively small, about 15
absorption system/absorbing galaxy pairs.  These studies have
not provided any clear systematic trends to elucidate the nature of
absorbing gas in relation to the galaxies.  The ``scatter'' in the
properties is simply too high for statisitcally significant trends to
be seen in the small sample.  If systematic kinematics do exists, it is
highly probable that additional ``parameters'' are at play that once
invoked can illucidate the systematics. Thus, it is also of supreme
importance to explore \hbox{Mg~{\sc ii}} kinematic connections with
the galaxy morphologies, the line of sight paths of the quasar light,
and the kinematics of the galaxies themselves.

In Figure~\ref{fig:collage} we show the galaxy morphologies and
\hbox{Mg~{\sc ii}} $\lambda 2796$ absorption profile kinematics as a
function of the quasar--galaxy impact parameter for 10 galaxies.  The
galaxy images have been obtained with WFPC-2/{\it HST\/} and the
\hbox{Mg~{\sc ii}} profiles with HIRES/Keck.  The images are $2\arcsec
\times 2\arcsec$; the galaxy impact parameters are given in the upper
portion of the panels and the redshifts in the lower portion.  The
absorbing gas is shown in the galaxy rest--frame velocity from $-220$
to $220$~km~s$^{-1}$ (the zero points are arbitrary).  The
\hbox{Mg~{\sc ii}} absorption exhibits a wide range of kinematic
complexity, with velocity spreads of $\sim 150$~km~s$^{-1}$, at
various impact parameters.  Ticks above the profiles give the numbers
and velocities of subcomponents, based upon $\chi ^{2}$ Voigt profile
fitting \citep{cv01,cvc03}.

We have used GIM2D \citep{simard} to model the galaxy morphologies for
26 absorbing galaxies \citep[see][]{glennk}. In
Figure~\ref{fig:glennfig}, we show, from top left to top right, the
{\it HST\/} image, the two--component GIM2D model, and the model
residuals.  , i.e., $\hbox{data} - \hbox{model}$.  North is oreinted
up and east to the left.  Note the striking asymmetric bar structure
and the ``tail'' extending to the southeast.  These structures are not
immediately apparent in the data prior to modeling.  We have found
that similar levels of asymmetric structures are present in many of
the galaxies in our sample.  That is, the galaxy morphologies appear
fairly normal prior to examination of the model residuals
\citep{steidel98,glennk}.

\begin{figure}[hbt]
\plotfiddle{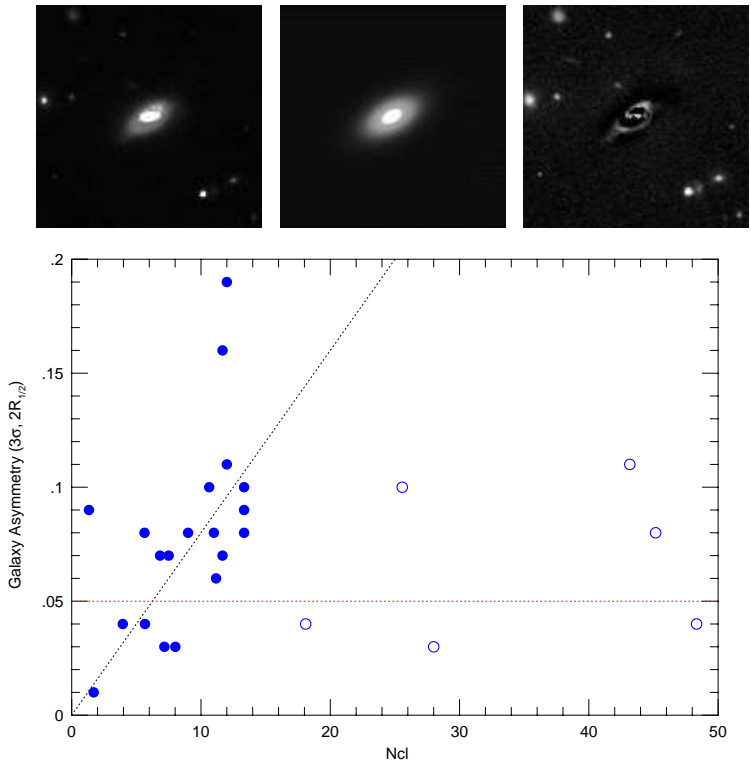}{3.8in}{0}{60.}{60.}{-160.}{-30.}
\caption{{\bf Morpholgy and Kinematic Spread} --- (top, left) The
WFPC-2/{\it HST\/} image of an absorbing galaxy quasar.  --- (top,
center) The GIM2D model of the galaxy. --- (top, right) The model
residual.  Note the asymmetric morphology. --- (bottom) The morphology
asymmetry versus the number of clouds in \hbox{Mg~{\sc ii}}
absorption.  The data suggest a correlation between morphological
``disturbance'' and qas quantity and/or gas kinematics.}
\label{fig:glennfig}
\vglue -0.1in
\end{figure}

For each galaxy in our sample, we have quantified the residual
asymmetries using the parameter $R_A$ \citep{schade95} for $3~\sigma$
residuals out to twice the half light radius, $R_{1/2}$.  When $R_A >
0.05$, the galaxy is considered to have significant morphological
assymetry (rotated $180^{\circ}$ about the galaxy axis).  In the lower
panel of Figure~\ref{fig:glennfig}, we plot $R_A$ versus $N_{cl}$, the
number of ``clouds'' (Voigt profile components) comprising the
\hbox{MgII~{\sc ii}} absorption profile.  Filled data points have
$W(\hbox{Mg~{\sc ii}}) \leq 1.0$~{\AA} and open data points have
$W(\hbox{Mg~{\sc ii}}) > 1.0$~{\AA}.  Note that 18 of the 26 galaxies
are catagorized as having significant asymmetric morphologies.  

We limit further discussion to those systems with $W(\hbox{Mg~{\sc
ii}}) \leq 1.0$~{\AA}. These are ``classical'' systems, whereas those
with larger $W(\hbox{Mg~{\sc ii}})$ typically are ``double'' systems;
those with $W(\hbox{Mg~{\sc ii}}) \geq 1.5$~{\AA} are typically DLAs
with severe profile saturation \citep{archiveII}.  There is a
$2.8~\sigma$ correlation between $R_A$ for the galaxy and $N_{cl}$ for
the absorbing gas.  The dotted line through the data is not a fit,
only a guide for the eye.

We also tested for correlations between gas kinematics and galaxy
orientation.  Orientation was parameterized by the angle between the
galaxy major axis and the quasar sightline, and by the inclination of
the galaxy.  Only three of our 26 galaxies are ellipticals.  We also
examined orientation normalized to the galaxy luminosity and the
quasar--galaxy impact parameter.  No significant correlations were
found.

\section{First Hints of a Galaxy--Halo Gas Kinematic Connection}

At least in our moderate sized sample, there is no clear indication
that the geometric cross section and kinematics of the gas is closely
related to orientation of the line of sight through the galaxy
``halo''.  However, the correlation between galaxy morphological
asymmetry and the number of clouds intercepted by the quasar sightline
is our first hint of a connection between the galaxy proporties and
the gaseous conditions in extended gas.  It suggests that
gravitational disturbance, possibly due to a previous merging or
accretion event, may have induced a stirring of gas surrounding the
galaxy.  Such disturbances are often seen in local galaxies from
21--cm studies \citep[see][{\it The Rogue Catalog}]{hibbard01}.  The
fact that, for our sample, the luminous morphologies of the galaxies
are fairly normal would imply that the events are not neccessarily
violent, or that they occured a few dynamical times prior to the epoch
of observation.

Ultimatley, we need to obtain the kinematics of the luminous
components of the galaxies.  \citet{steidel02} recently conducted a
detailed study comparing the galaxy kinematics to the gas kinematics
for five \hbox{Mg~{\sc ii}} absorbers.  We caution that these galaxies
are all edge--on spirals, and not representative of the overall
population of \hbox{Mg~{\sc ii}} absorption--selected galaxies.
For four of the five, the absorbing gas
kinematics is rotating in the same sense as the galaxy; however, the
total spread of gas velocities is inconsistent with simple disk
rotation.  In three of the five cases, the gas kinematics can be
explained if the halo rotation ``lags'' the galaxy disk rotation, as
is seen with DIGs and ``beards'' in local galaxies
\citep{swater97,rand00}, but we are probing much higher in the halo.
As a caveat, we offer a counter example: \citet{ellison03} showed that
the \hbox{Mg~{\sc ii}} gas kinematics in an edge--on spiral galaxy
toward MC~$1331+170$ does not {\it fully\/} follow the disk rotation.
Some of the gas is akin to the aforementioned ``forbidden'' gas,
moving opposite to the rotation \citep{fraternali02}.

\end{document}